\documentstyle[12pt]{article}
\begin{document}
\baselineskip=24pt

\begin{titlepage}

\begin{center}
{\bf TOWARDS THE NON-CHIRAL EXTENSION OF SM AND MSSM}
\end{center}

\vspace{1 cm}

\centerline {J.L.~Chkareuli\footnote {E-mail:
jlc@physics.iberiapac.ge, jlc@gps.org.ge}, I.G.~Gogoladze~ and~
A.B.~Kobakhidze }

\centerline {\sl Institute of Physics, Georgian Academy of Sciences, 380077
Tbilisi, Georgia }

\begin{abstract}
We show that in some SU(N) type GUTs with the complementary pairs
of the conjugated fermion multiplets there naturally appear the
relatively light ($M\ll M_{GUT}$) vectorlike fermions which
considerably modify the desert physics.  In the non-SUSY case they
can provide for the unification of the standard coupling constant
$\alpha_1$, $\alpha_2$ and $\alpha_S$ whereas in the SUSY case they
can increase the unification point up to the string unification
limit and decrease $\alpha_S(M_Z)$ down to the value predicted from
the low energy physics.
\end{abstract}

\vspace{1 cm}

\end{titlepage}

\section{Introduction}

The LEP confirmation \cite{1} of three neutrino species in Z-bozon
decays seems to mean that we have in general only three standard chiral
quark-lepton families. Actually, there is not in fact a viable way to the
heavy ($m_{\nu}>M_Z/2$) neutrino for the hypothetical fourth family
in the framework of the SM to say nothing of the GUTs. So, one
can say after observation of the top quark at FERMILAB \cite{2} that all
the chiral matter of the SM are already discovered and now only its
SUSY counterpart, "smatter" of the MSSM, remains to be detected.
Needless to stress specially that not only the MSSM but the ordinary
minimal GUTs like as SU(5) or SO(10) \cite{3} and their SUSY
versions \cite{4} also suggest the pure chiral extension of the SM.

What one could say now about the fermion matter-fields vectorlike under the
SM, $\Psi _{SM} + \overline {\Psi}_{SM}$? If they exist, where could
be their mass scale? The common experience tells us that it could be
somewhere near the scale of the "next" chiral symmetry manifesting itself
generally in the form of $SM\otimes D$ or $GUT\otimes D$ or possibly
$D\supset GUT$ where D stands for discrete or global or even local
symmetry covering the known minimal GUTs SU(5) or SO(10) (the letter
abbreviations such as SM, GUT etc. are used everywhere for the underlying
symmetries as well). In contrast to the SM D-symmetry somehow
differentiates the left- and right-handed components of the above
vectorlike matter-fields and thus protect their masses from being much
heavier than its own scale $V_D$.  The "next" chiral symmetry could
be some family symmetry H (say, a chiral $SU(3)_H$ symmetry
acting in 3-dimensional generation space \cite{5}) or Peccei-Quinn
symmetry $U(1)_{PQ}$ \cite{6} both concerning besides the ordinary SM
chiral quarks and leptons some vectorlike pairs of the fermion
multiplets conjugated under the SM. These multiplets receive their
masses of order $V_H$ or $V_{PQ}$ after H-symmetry or $U(1)_{PQ}$ is
spontaneously broken. One more example could be suggested by
non-minimal GUTs, e.g.  $SU(N>5)$ \cite{7} which contain generally
many vectorlike fermion pairs ($5+\overline 5$) and ($10+\overline
10$) of SU(5). In the ordinary exposition \cite{7} their masses are
appeared to be about the scale of breaking  of the SU(N) GUT down to
SU(5) and thus have no real influence even on the near-GUT physics
not to mention the low-energy one.  However, their could be the other
breaking channels as well (not necessarily following through the
Georgi-Glashow SU(5) \cite{3}) giving lower masses for the non-chiral
fermions \cite{8}. So, to conclude some additional fermion matter, if
it exists, should certainly be vectorlike under SM or even GUT and be
accommodated in general somewhere in the grand desert between the SM
and GUT scales.

Another question is do we really need any additional fermion matter beyond
the SM multiplets of quarks and leptons. The answer could be positive
if we wanted to overcome the crisis, related with the actual
non-unification of the standard coupling constants $\alpha_1$, $\alpha_2$
and $\alpha_S$ in the SM \cite{9}. Also in the MSSM this unification
\cite{9} does not look enough conclusive if one keep in mind the real gap
between the SUSY GUT and string unification cases \cite{10} as well as
some discrepancy for $\alpha_S(M_Z)$ predicted from the single scale SUSY
GUT unification on the one hand and low energy data on the other
\cite{11}. It has been shown recently \cite{11} that in the minimal SU(5)
type theories, the inclusion of the threshold corrections does not change
this situation.  Thus one could expect that some new physics might appear
in the grand desert accommodating the above SM vectorlike matter at the
certain intermediate scale $V_D$ so that to improve the running of the
standard coupling constants correspondingly.

In this letter we show that along the certain breaking channels of the
$SU(N)\otimes D$ type GUTs extended to include the additional complementary
pairs of the SU(N) conjugated fermion multiplets there naturally appear the
relatively light ($M\ll M_{GUT}$) vectorlike fermions depending on a group
order and starting multiplets involved. They turn out to considerably
modify the desert physics.  In the non-SUSY case they provide for the
unification of the standard coupling constants, whereas in the SUSY case
they can increase the unification point up to the string unification limit
and decrease $\alpha_S(M_Z)$ down to the value predicted from low-energy
physics.

\section{Hyperneutral split fermions}

So, we start with a general $SU(N)\otimes D$ GUT containing besides some
"standard" anomaly free set of fermion multiplet with ordinary quarks and
leptons
\begin{eqnarray}
3\cdot \biggl [(N-4)\cdot \overline {\psi}~
{N \choose 1} + \psi~
{N \choose 2}
\biggr ]
\label{1a}
\end{eqnarray}
the $n_F$ pairs of the conjugated chiral fermions (complementary
fermions)
\begin{eqnarray} n_F\cdot \biggl[\Psi _1~{N \choose K}
 + \overline {\Psi} _2~
{N \choose K}
\biggr ]~~,
\label{1b}
\end{eqnarray}
(in the left-handed basis) where ${N \choose K}$ stands for dimension
of the particular SU(N) antisymmetric representation used merely to
have only the quark and lepton type states in $\Psi _1$ and
$\overline {\Psi}_2$.

The most natural genesis of the SM from the general SU(N) theory looks
as the spontaneous breakdown of the starting $SU(N)\otimes D$ symmetry due
to the "standard" scalar set which includes one adjoint ($\Phi ^i_j$) and
$N-5$ fundamental ($\varphi _i^{(r)}$) heavy scalars of SU(N) ($i,j =
1,...  N; r = 1,...N-5$)
\begin{equation}
SU(N)\otimes D \stackrel{\Phi}{\longrightarrow}
SU(n)_S\otimes SU(N-n)_W\otimes U(I)
\stackrel{\varphi ^{(r)}}{\longrightarrow} SU(3)_S\otimes
SU(2)_W\otimes U(1)
\label{2}
\end{equation}
accommodating generally the
strong and weak parts of the SM in the different subgroups $SU(n)_S$ and
$SU(N-n)_W$, respectively. The natural case when the scalars $\Phi$ and
$\varphi ^{(r)}$ have the same order VEVs ($\Lambda \sim \lambda ^{(r)}$)
corresponds to the minimal (one-scale) SU(N) GUTs broken down to the SM
below the unification point. It goes without saying that besides $\Phi$ and
$\varphi ^{(r)}$ scalars there are generally two scalar field
multiplets $H_1$ and $H_2$ which breaks subsequently the SM and give
masses to the $up$ and $down$ quark, respectively.

The simplest choice for chiral D symmetry here in the framework SU(N) GUTs
seems to be the familiar reflection \cite{4} for the adjoint scalar $\Phi$
accompanied now by appropriate reflection in the vectorlike pairs (2)
\begin{equation}
\Phi \rightarrow -\Phi,~~ \Psi _1 \rightarrow \Psi _1,~~
\overline {\Psi}_2 \rightarrow -\overline {\Psi}_2
\label{3}
\end{equation}
so that their direct $SU(N)$ invariant mass term is fully suppressed and only
their Yukawa couplings with scalar $\Phi ^i_j = (\phi ^A\cdot T^A)^i_j$
\begin{equation}
G\cdot \overline {\Psi}_2\Phi \Psi _1 + h.c.
\label{4}
\end{equation}
($G$ is a coupling constant and $T^A$ are generators of $SU(N)$) are allowed
to exist. So, the masses of the vectorlike fermions (2) will
completely determined by the VEV matrix $\langle \Phi ^i_j \rangle$
only. It is well known \cite{12} that the adjoint scalar $\Phi $ itself
develops VEV along one of the hypercharges $\hat {Y}_N^{(n,N-n)}$ of
SU(N) providing for the first stage in the breaking chain (3). What
this means is the submultiplets in $\Psi _1$ and $\overline {\Psi}_2$
which have $Y_N=0$ drop out of the basic coupling (5) and thus leave
to be massless until the fundamental scalars $\varphi ^{(r)}$ break $Y_N$
making the adjoint scalar $\Phi $ develop the VEV along the other
hypercharges as well. Actually, the fundamental scalars $\varphi ^{(r)}$
having no the direct Yukawa couplings with $\Psi _1$ and $\overline
{\Psi}_2$ affect their mass spectrum only through the intersecting terms
in the general Higgs potential $V$ of scalars $\Phi $ and $\varphi ^{(r)}$
\begin{equation}
V(\Phi, \varphi ^{(r)}) = ...+ a(Tr\Phi ^2)^2 + b(Tr\Phi ^4) +
\alpha \overline {\varphi}\varphi Tr\Phi ^2 + \beta \overline
{\varphi}\Phi ^2\varphi + ...
\label{5}
\end{equation}
(two last terms in (6), index $r$ is omitted) inducing in the
starting VEV matrix $\langle \Phi ^i_j \rangle$
\begin{equation}
\langle \Phi ^i_j \rangle = \Lambda diag\biggl [ 1...1,
-\frac{n}{N-n},...-\frac{n}{N-n} \biggr ]^i_j =
\Lambda\biggl [ {Y}_N^{n,N-n} \biggr ]^i_j
\label{6}
\end{equation}
some other the SM invariant corrections of order
\begin{equation}
\epsilon ^{(r)}\Lambda,~~~~~ \epsilon ^{(r)} = \frac{\beta
^{(r)}}{b}\biggl (\frac{\lambda ^{(r)}}{\Lambda}\biggr )^2
\label{7}
\end{equation}
during the second stage of the symmetry breaking process
(3). In such a manner the mass matrix of the complementary fermions
(2) can be expressed generally through the hypercharges of $SU(N)$ as
\begin{equation} \hat M = M_0\cdot \hat Y_N + \sum_{r=1}^{N-5}M_s\cdot \hat
Y_{N-r},~~ M_0 = G\Lambda, M_r = G\Lambda \epsilon ^{(r)}
\label{8}
\end{equation}
(correct to some calculable group factors in $M_0$ and
$M_r$), where $\hat Y_N$ corresponds to the hypercharge matrix of
U(I) in (3) and the others ($\hat Y_{N-r}$) belong to $SU(N-r)$
groups while the last one $\hat Y_5$ is the familiar hypercharge of
the standard SU(5) \cite{4} or what is the same the normalized
hypercharge of the SM.

One can see now that we are driven at the natural mass-splitting inside of
the fermion pairs (2) depending on the U(I) hypercharge values of
their $SU(n)_S\otimes SU(N-n)_W$ submultiplets. While the general mass
scale of the $\Psi$ pairs (2) is given by the largest mass $M_0$ in Eq.(9)
their U(I) neutral submultiplets survive the first stage of the symmetry
breaking in (3) and receive masses of the order of $M_r$ only during
subsequent breakings down to the SM. This mass gap in the split $\Psi$
multiplets (2), even within the soft radiative hierarchy between VEVs
$\lambda ^{(r)}$ and $\Lambda$ or, preferably, between coupling constants
$\beta ^{(r)}$ and $b$ in Eq.(8), appears to strongly affect the running
of the standard gauge coupling constants $\alpha _1, \alpha _2$ and $\alpha
_S$ in the superhigh energy area (see below).  Some motivation for $\beta
^{(r)}\ll b$ would stem from that the corresponding coupling in the Higgs
potential (6) has the starting Lagrangian (local) symmetry SU(N) only,
whereas the other terms in it are invariant under independent
$SU(N)_{\Phi}$ and $SU(N)_{\varphi}$ transformations as well. Thus the last
term in the potential (6) could purely be induced by the $SU(N)$ gauge
loops giving the natural radiative order for constant $\beta$,~~ $\beta
\sim \alpha ^2_{GUT}$.  So, even in the case of the single point
unification ($\Lambda \sim \lambda ^{(r)}$) there would be a few order
hierarchy between the mass parameters $M_0$ and $M_r$ ($\frac{M_0}{M_r}\sim
\alpha ^2_{GUT} = 10^3-10^4$) in Eq.(9)\footnote {Of course, this hierarchy
could be considerably enhanced if we suppressed a constant $\beta$ by some
special fine tuning or the SUSY arguments (see below).}. It seems to be
quite remarkable that while in that case the SU(N) GUT breaks down to the
SM at once, the $\Psi$ multiplet mass spectrum still follow to the
two-step breaking process (3) generating among others the relatively
light masses of the U(I) neutral submultiplets in $\Psi$. So, the
reflection (4) not only protect the complementary fermions from
having the heavy SU(N) invariant masses but also provide for some
their submultiplets the masses much lower then its own scale
$\Lambda$. We call them the hyperneutral split fermions (HSF).

\section{The HSF scanning the SU(N) GUTs}

Let us imagine for a moment that we know nothing about the SU(N) GUTs at
all and will scan them now depending on the mass spectrum of the
complementary fermions only - whether there are allowed to exist the
hyperneutral split fermions or not.

Let the complementary fermion multiplets $\Psi _1$ and $\overline {\Psi}_2$
(2) belong to some pure antisymmetric representation of SU(N) broken
to $SU(n)_S\otimes SU(N-n)_W\otimes U(I)$. One can see that the HSF
submultiplets will appear if the following group conditions are satisfied
\begin{equation}
\frac{k_n}{n} = \frac{k_{N-n}}{N-n} = \frac{K}{N}
\label{9}
\end{equation}
where K is the order (number of indices) of the multiplets $\Psi _1$ and
$\overline {\Psi} _2$ under SU(N) whereas $k_n$ and $k_{N-n}$ are
suborders of their HSF submultiplets under subgroups $SU(n)_S$ and
$SU(N-n)_W$, respectively ($k_n + k_{N-n} = K$). Eq.(10) follows, by
definition, from the zero U(I) hypercharge value for the HSF
submultiplets as it can be derived by direct application the $\hat Y_N$
matrix (7) to them.

Now using the basic condition (10) we are able to carry out the general
classification of all the possible SU(N) GUTs depending upon the order
value K only of the complementary fermions:
\begin{quote}
(i)~ $K = 1$
does not lead to any HSF submultiplets. The only solution $n = N$ of
Eq.(10) corresponds to the non-broken $SU(N)$;

(ii)~ $K = 2$ gives $k_n = 2 {\frac{n}{N}} = 1$ (keeping in mind that
$k_n\leq K$ by definition, whereas $k_n = K$ conforms again with the
non-broken symmetry case) and leads to the SU(2n) GUTs with breaking
pattern $SU(n)_S\otimes SU(n)_W\otimes U(I)$ and the HSF submultiplets in
the representation $(n,n) + (\bar n,\bar n)$ whose decomposition under SM
gives \begin{equation} \biggl [(3_c,2) + (3_c,1)_d + h.c.\biggr ] +
(n-3)\biggl [5 + \bar 5\biggr ]
\label{10}
\end{equation} apart from the
trivial SM singlets. Here index $d$ means the down quark type state in
$(3_c,1)$ and $(\bar 3_c,1)$ while $5$ and $\bar 5$ stand for the
standard SU(5) quintet and antiquintet, respectively.

(iii)~ In the general case $K = 1,2,...,\frac{N}{2} (\frac{N-1}{2})$ we are
led to the $SU(n + n\frac{k_{N-n}}{k_n})$ GUTs ($n = 1,...,N; k_n =
1,...,K; k_n + k_{N-n} = K$) with the breaking pattern $SU(n)_S\otimes
SU(n\frac{k_{N-n}}{k_n})_w\otimes U(I)$ and the HSF submultiplets in the
representation
\begin{eqnarray}
\biggl [~
{n \choose k_n}
~~,~~
{n\frac{k_{N-n}}{k_n} \choose k_{N-n}}
~ \biggr ] + h.c.
\label{11}
\end{eqnarray}
from where the previous particular cases can easily be reproduced.
\end{quote}

So, keeping in mind $N\geq 5$ for the SU(N) GUT covering the SM we
should conclude that a familiar SU(5) can not satisfy the criterion
condition (10) for any non-trivial HSF submultiplet and has be
excluded completely. The same occurs for the other prime order SU(N) GUTs
$(N = 7,11,...)$ unless as they preliminary break to one of the allowable
cases (see (iii)) due to the some other mechanism and thereupon follow to
the our HSF scenario.

The another point is that among the all above HSF versions the case (ii)
seems to be certainly singled out. As one can see from Eq.(11) there
appear that all SU(2n) theories contain in their HSF sectors the common SM
fragments plus the SU(5) full quintets which does not affect (in the
leading 1-loop order) the unification picture. Thus this picture could be
expected very similar for the whole class of the SU(2n) GUTs as we found
there the practically order-invariant HSF submultiplets.

An inspection of this order-invariant part of the HSF submultiplets (the
first term in Eq.(11)) together with two\footnote{One can see that all
the GUTs beyond the minimal SU(5) require at least two independent
scalar doublets for the $up$ and $down$ quark sectors, respectively.}
electroweak doublets in the starting Higgs field $H^1$ and $H^2$ in the
model shows that we have directly driven just at so called ABC
split-multiplet Anzatz postulated in the framework of the standard SU(5) a
decade ago by Glashow and Frampton \cite{13} and thoroughly revised
recently by Amaldi et al \cite{9} after they have examined over $1600$
split-multiplet combinations of quarks, leptons and scalars.

The ABC model is known \cite{9} to be well consistent with present data
and give perfect-single point unification when light split fermions are
taken on the low mass scale near the TeV region while their heavy partners
are on the grand one. It is easily comprehended on the other hand that such
a situation requires a new special fine-tuning between the gigantic VEVs
of the Higgs scalar in $1, 24$ and $75$ reps of the SU(5) to get in
general the rather light split-multiplet fermions in the SU(5) model.
Thus "staying alive with SU(5)" \cite{13} seems to be even more
problematic than the old hierarchy problem.

However, it seems to be quite reasonable to think that if one family of the
split-multiplet fermions ($n_F=1$ in Eq.(2)) have to start rather early
to correct a right way the running of the constants $\alpha _1,
\alpha _2$, and  $\alpha _s$ two or three families of them could start
later to do the same. Thus, we could expect that instead of one light
family there appear two or would be better three (as for ordinary quarks
and leptons) heavy families of split-multiplet fermions. Ideally, their
mass scale $M_{HSF}$ could be arranged at the "radiative distance" from the
grand scale $M_G$, $M_{HSF}\sim \alpha ^2_{GUT}M_G$ so as not to have above
mentioned hierarchy problem for the HSF spectrum. Fortunately, it happens
to be the case in our model (see Table 1).

So, there appear at first time not only to derive the ABC Anzatz
theoretically with our HSF scenario in the framework of the general SU(2n)
GUTs but also to avoid the split fermion mass hierarchy problem
introducing the several families of the HSF states. Considering the minimal
possible GUT (n=3) we are led to SU(6) model with total fermion content
(1,2) as
\begin{equation}
3\cdot (~2\cdot \bar 6 + 15) + n_F\cdot (15 + \overline {15})
\label{12}
\end{equation}
containing only order-invariant part in Eq.(11).

Using then as input parameters the World average values $\alpha _s(M_Z) =
0.117\pm 0.005$, $\alpha _{EM}(M_Z) = 1/(127.9\pm 0.2)$ and $sin^2\theta
_W(M_Z) = 0.2319\pm 0.0008$ \cite{14} in the standard RG equations for
the running constant $\alpha _1, \alpha _2$, and  $\alpha _s$
\begin{equation} \mu\frac{d}{d\mu}\alpha^{-1}_{i}=-\frac{1}{2\pi}
\biggl(b_i+\frac{b_{ij}}{4\pi }\alpha_j+O(\alpha^{2}_i)\biggr)
\label{13}
\end{equation}
with the 1-loop and 2-loop $b$-factors [15,9] we are driven at Table 1
showing a perfect single-point unification for different values of family
number $n_F$ of HSF states. Clearly, this picture will practically hold
true in general case SU(2n) GUT as well.

Although, we used hitherto the electroweak angle value as input parameter
to present the extended picture possible the model allows to predict in
principle this angle itself if we start with three (or four) families of
HSF states and consider the mass scale $M_{HSF}$ of split fermions as of
the pure radiative origin ($M_{HSF}\sim \alpha ^2_{GUT}M_G$). So, the
SU(2n) GUTs with three families of the complementary fermions seems to
compare well with MSSM as to observable aspects of unification. However,
the SUSY extension of the above SU(2n) theories are of special
interest and we are coming to it now.

\section{The HSF room in the SUSY SU(2n) GUTs}

Let us consider the minimal SU(6) model keeping in mind the whole
class of the SU(2n) GUTs. The essential point related with the SUSY
extension of those GUTs seems to be that they break themselves mainly
along the foregoing $SU(n)_S\otimes SU(n)_W\otimes U(I)$ channel
providing thus the natural room for the HSF submultiplets.

The most general superpotential for the above heavy fields (chiral
superfields now) $\Phi ^{i}_{j}$ and $\varphi _{i}(\overline {\varphi}^{i})$
in our SU(6) model looks as
\begin{equation}
W = \frac{\mu}{2}Tr\Phi ^2 + \frac{h}{3}Tr\Phi ^3 + m\overline {\varphi}
\varphi + \lambda\overline {\varphi}\Phi \varphi
\label{14}
\end{equation}
(the light Higgs supermultiplets $H_i(\overline {H}^i)$ and
$H_{ij}(\overline {H}^{ij})$ of SU(6) generating masses of the $down$ and
$up$ quarks (1), respectively, are not essential for the present
discussion). The standard supersymmetric analysis of the F-terms of the
heavy superfields $\Phi$, $\varphi$ and $\overline \varphi$
\begin{equation}
F_{\Phi}= F_{\varphi}=F_{\overline {\varphi}}=0
\label{15}
\end{equation}
during breaking process (3) shows that the
trivial non-broken case apart the only symmetry breaking pattern of
SU(6) in the no-scale limit of a superpotential $W_0=W(\mu=m=0)$ is just
the HSF channel $SU(3)\otimes SU(3)\otimes U(I)$
\begin{equation}
\langle
\Phi _j^{i} \rangle  = \Lambda diag[1,1,1,-1,-1,-1]_j^i~~~~ \langle \varphi
\rangle = \langle \overline {\varphi} \rangle =  0 \label{16}
\end{equation} where the VEV parameter $\Lambda$ is not fixed as yet. The
switching on the masses $\mu$ and $m$ in $W$ (15) does open the other
channels as well leading among all the degenerate vacua to the familiar
SM one with the VEVs of the scalars as \begin{eqnarray} \langle \Phi
_j^{i} \rangle  = \frac{m}{\lambda} diag[1,1,1,-1,-1,-1]_j^i +
\frac{\mu}{h} diag[2,2,2,-3,-3,0]_j^i  \\ \nonumber \langle \varphi _i
\rangle = \langle \overline {\varphi}^i \rangle = \biggl [ 6
\frac{\mu}{\lambda}(\frac{m}{\lambda} + \frac{\mu}{h})\biggr ]^{1/2} \delta
_{i6} \label{17} \end{eqnarray} An interesting feature of the solution (17)
seems to be that, while the supposedly largest rank-preserving part in the
$\Phi$ scalar VEV are determined by the mass parameter $m$ of the
fundamental scalar $\varphi$, in its own VEV $\langle \varphi\rangle$ the
contributions of the $O(m)$ order are cancelled and the leading order is
just the average geometrical one $O(\sqrt {\mu m}),~~ m\gg \mu$. So, there
is an automatic $SU(3)\otimes SU(3)\otimes U(I)$ intermediate gauge scale
$M_I= O(\sqrt {\mu m})$ in this case by contrast to the non-SUSY one
(Sec.3) where we have (inspite of the dynamically appearing HSF scale)
the single point gauge unification. However, as we can see below, this new
scale is turned out also to be related with the HSF submultiplets.

The above SUSY analysis of the SU(6) symmetry breaking pattern remains in
force after a typical low-energy SUSY breaking as well. The standard
effective Higgs potential following from the minimal $N=1$ Supergravity
\cite{16}
\begin{equation}
V=\biggl
|\frac{\partial W}{\partial z_i} + m_{3/2}z^{*}_{i}\biggr |^2 +
m_{3/2}(A-3)[W^{*} + W] + D-terms
\label{18}
\end{equation}
can induce only the little shifts in the VEV in Eq.(18).of the gravitino
masses order $O(m_{3/2})$ at most. At the same time Supergravity could
provide some reasoning for lifting vacuum degeneracy in favor of just the
minimum (18) in the general case as well.

Now coming back to the supersymmetric analogue $L^{SUSY}_Y$ of the Yukawa
coupling (5), which continues to be invariant \footnote{We could say that
this invariance holds in the no-scale superpotential $W_0$ as well in form
of the $R$-parity $$\Phi\rightarrow -\Phi,~\varphi(\overline
\varphi)\rightarrow -\varphi (\overline
\varphi),~W_0=-W_0,~L_Y^{SUSY}=-L_Y^{SUSY}$$ what leads to the pure
HSF vacuum (17).  The mass terms in the general superpotential W (15)
break this symmetry and induce the "scaled" solution (18) among the
other degenerate ones.} under reflection (4) suppressing the direct
mass term for the complementary fermions (and sfermions) (2), we are
led from the VEV matrix $\langle \Phi ^i_j\rangle$ (16) to the above
mass formula (9) adapted to our SU(6) case \begin{equation} \hat M =
M_6\hat Y_6 + M_5\hat Y_5,~~~M_6 = G\frac{m}{\lambda}\sqrt{12},~~ M_5
= G\frac{\mu}{h}\sqrt{60} \label{19} \end{equation} where $G$ is
Yukawa coupling constant and factors $\sqrt{12}$ and $\sqrt{60}$
appeared from the standard normalization of hypercharges $Y_6$ and
$Y_5$, respectively. It can be easily seen now that any gap between
masses $\mu$ and $m$ of the scalars $\Phi$ and $\varphi$ are
immediately transformed into the gap inside of the complementary
particles in the rep $15 + \overline {15}$ of SU(6) (see Eq.(13))
where the SUSY HSF submultiplets $(~3.3~) + (~\bar 3. \bar 3~)$ have
relatively light masses ($M_{HSF} = \frac{\mu}{h}G$) while the
remainders $(~3.1~) + (~\bar 3. 1~) + (~1.3~) + (~1.\bar 3~)$ are
much heavier ($M_F = \frac{m}{\lambda}G = M_GG$, $M_G$ is the
unification mass) up to the grand scale ($G\simeq 1$). According to
Eq.(18) and (20) the $SU(3)\otimes SU(3)\otimes U(I)$ intermediate
scale $M_I$ given by the VEV of the scalar $\varphi (\overline
{\varphi})$ can be expressed now through the basic parameters of the
model - the unification mass $M_G$ and the HSF scale $M_{HSF}$
\begin{equation} M_I = [M_GM_{HSF}]^{1/2}\cdot \eta,~~~\eta = \biggl
(6\frac{h}{G\lambda}\biggr )^{1/2} \label{20} \end{equation} While in
the ordinary case (Sec.3) the natural gap between masses $M_{HSF}$
and $M_G$ would be at most the radiative one now in the $SUSY$ case
mass $M_{HSF}$ could be in principal anywhere below and even down to
the SUSY scale.

One can understand, that in the framework of the ordinary MSSM giving the
perfect single point unification \cite{9} it would look quite hopeless to
find any HSF states somewhere beyond the unification area itself.
However, in our case of the dynamically stipulated intermediate unification
with the calculable scale (21) such a possibility appears even for the
standard low-energy SUSY breaking strongly correlated with the
electroweak scale, say $M_{SUSY} = 3.2\cdot 10^2 GeV$. Our results are
presented in Table 2 for one family of the HSF states and three different
values of $\alpha _s(M_Z)$ after the above RG equations (14) (with $b_i$
and $b_{ij}$ factors including contributions of the supersymmetric partners
as well \cite{15}) were numerically integrated in 2-loop level approximation.
At the same time it seems to be of a special interest the question about
the SUSY scale itself as following from the unification requirement only.
The general dependence of the unification scale $M_G$ (dashed line) and the
HSF scale $M_{HSF}$ (solid line) from the SUSY scale $M_{SUSY}$ for the
central value $\alpha _s(M_Z)=0.117$ are given on Fig.1 for one HSF family
again (in $\eta=1$ case, see Eq.(21)). One can see clearly that the
under-Plank mass unification requirement ($M_G\leq M_{Pl}$)and general
condition $M_{HSF}\geq M_{SUSY}$ following from the effective $N=1$
Supergravity potential (19) leave the only possible area for the SUSY scale
$M_{SUSY} = 3\cdot 10^2\div 10^3 GeV$ and correspondingly for HSF scale
$M_{HSF} = 10^{11}\div 10^{16} GeV$. The possible though not too
restrictive limitation on the HSF room in this model could be expected from
the $b-\tau$ unification, stipulated by ordinary Yukawa couplings of quarks
and leptons with the Higgs supermultiplets $H_i(\overline H^j)$ and
$H_{ij}(\overline H^{ij})$ (see above). While we are going to discuss it
closely in a separate publication we include in the Table 2 some b-quark
mass 1-loop values $m_b(m_b)$ taking in its RG equation the "maximal" top
quark Yukawa constant on the unification scale $Y_t(M_G)=1$
($Y_t=G_t^2/a\pi$) and as top quark mass value $m_t=174 GeV$. One can see
that for the parameters involved somewhat large value of $m_b$ appears for
the Plank mass unification only.

So, we are driven at a conclusion that there could be some new physics in
$10^{11} - 10^{14} GeV$ region related with non-chiral extension of the
MSSM. The above HSF particles affect the unification considerably
increasing its scale to the string unification limit $M_{Str}\approx
5\cdot 10^{17} GeV$ and even up to Plank mass $M_{Pl}$.
Simultaneously, they could lead to rather low value $\alpha _s(M_Z)$
predicted from the low-energy physics \cite{11} (see Table 2). In
contrast to the non-SUSY case the SUSY SU(6) (and SU(2n) GUTs in
general) strongly prefer one HSF family case accommodating multi-family HSF
states in the vicinity of the unification area itself ($M_{HSF}\approx
M_{I}\approx M_G$).

\section{Summary}

We have discussed as general as possible the problem of inclusion of the
non-chiral matter in the SM and MSSM and found that there could naturally
exist the special set of the relatively light HSF particles in the
framework of the SU(2n) type GUTs. So far we knew only two "canonical"
sets of the particles which ruled the unification phenomena in GUTs - the
ordinary SM set with quarks, leptons and Higgs doublets, and the MSSM set
including all their supersymmetric partners as well. The first set itself
is turned out no to be enough to give the unification at all. Also the
second set, the MSSM, while giving a perfect unification at $M_G\approx
10^{16} GeV$ seems not to be enough to give the very desirable higher
string unification at $M_{Str}\approx 5\cdot 10^{17} GeV$.

The third set, derived here from the starting fermion spectrum (1,2) is the
SM set plus HSF states (11), gives the perfect single point unification
just at the same point $M\approx 10^{16} GeV$ (see Table 1) as in the MSSM
case. Thus, the HSF states works exactly like as SUSY partners of the SM
particles for the running constants $\alpha _1$, $\alpha _2$ and $\alpha
_S$. Lastly, the fourth set which is just the SUSY version of the third one
(SM + HSF) leeds to the higher string and even Plank scale unification
depending on the mass spectra of the HSF states. It seems to be quite
interesting that in the both cases - ordinary and SUSY - HSF states
naturally happen to be on the rather higher mass scale $10^{11} - 10^{14}
GeV$ (see Secs.3 and 4) and considerably modify the desert physics.

While at present there are no any direct indication in favor of the SU(2n)
GUTs some arguments look to be relevant:  \begin{quote}

(i)~ In contrast to the MSSM \cite{11} strong coupling constant $\alpha
_S(M_Z)$ extrapolated down from the unification scale meets the value
extracted from the low-energy physics for ordinary (Sec.3, Table 1) as well
as SUSY (Sec.4, Table 2) cases;

(ii)~ In the higher SU(2n) symmetry cases SU(8), SU(10) etc. containing
gauge family symmetries there could appear among the others the
non-suppressed flavour-changing proton decay modes like as $p\rightarrow
\pi ^{0}\mu ^{+}, K^{0}e^{+},...$

(iii)~ An introducing of the complementary matter multiplets (2) in
addition to the ordinary one of the SM or MSSM could help to resolve
familiar vacuum $\theta$-domain problem \cite{17} in a manner by Georgi and
Wise \cite{18}.  \end{quote} We will consider all those and related
problems elsewhere.

\vspace{0.5 cm}

{\bf Acknowledgements}

We thank Z.Berezhiani and G.Dvali for interesting discussions. One of
us (J.L.C.) is deeply grateful to R.Barbieri, S.Dimopoulos,
H.Leutwyler, P.Minkowski and G.Senjanovich for stimulating conversations
and G.Veneziano for warm hospitality at CERN Theory Division where the part
of this work was made.

We acknowledge partial the  support of the International Science Foundation
under the Grant No.MXL000.

\newpage
\pagestyle{empty}

{\bf Table 1}~~The values of the split fermion scale $M_{HSF}$,
unification scale $M_G$ and the inversed unified constant $\alpha
^{-1}_{GUT}$ for different number of the HSF families ($n_F=1,2,3$)
depending on $\alpha _s=0.117\pm 0.005$.

\vspace{2 cm}

\begin{center}
\begin{tabular}{|c|c|c|c|}
\hline
& & & \\
$n_F$ & $M_{HSF}$, GeV & $M_{G}$, GeV & $\alpha ^{-1}_{GUT}$ \\
& & & \\
\hline
& & & \\
1 & $4.6^{-3.9}_{+30.1}\cdot 10^3$ &
$9.1^{+7.1}_{-4.2}\cdot 10^{15}$ &
$35.4^{-0.6}_{+0.6}$ \\
& & & \\
\hline
& & & \\
2 & $6.0^{-2.5}_{+6.9}\cdot 10^9$ &
$8.5^{+5.6}_{-3.9}\cdot 10^{15}$ &
$35.5^{-0.5}_{+0.5}$ \\
& & & \\
\hline
& & & \\
3 & $7.1^{-1.6}_{+8.9}\cdot 10^{11}$ &
$7.9^{+5.9}_{-3.4}\cdot 10^{15}$ &
$35.5^{-0.5}_{+0.5}$ \\
& & & \\
\hline
\end{tabular}
\end{center}

\newpage

{\bf Table 2}~~The values of the split supermultiplets scale
$M_{HSF}$, unification scale $M_G$ and the in inversed unified
coupling constant $\alpha ^{-1}_{GUT}$ for one family HSF states
depending on SUSY scale $M_{SUSY}=10^{2.5} GeV$ and different values
of $\alpha _s(M_Z)$. The intermediate scale $M_I$ is calculated to be
$M_{I}=10^{15.2}$ when parameter $\eta$ (see Eq.(21)) runs in some
natural region $0.5\div 1.5$. The 1-loop of b-quark mass values
presented for "maximal" value of top quark Yukawa constant $Y_t(M_G)=1
(m_t=174 GeV)$.

\vspace{2 cm}

\begin{center}
\begin{tabular}{|c|c|c|c|c|}
\hline
& & & & \\
$\alpha _s(M_Z)$ & $M_{HSF}$, GeV & $M_{G}$, GeV & $\alpha
^{-1}_{GUT}$ & $m_b(m_b)$, GeV \\
& & & & \\
\hline
& & & & \\
0.112 & $10^{13.6}$ & $10^{17.4}$ &
19.2&4.32  \\
& & & & \\
\hline
& & & & \\
0.117 &
$10^{12.0}$ & $10^{18.4}$ & 15.6 & 5.17 \\
& & & & \\
\hline
& & & & \\
0.122 & $10^{10.7}$ & $10^{19.2}$ & 12.5 & 5.66\\
& & & & \\
\hline
\end{tabular}
\end{center}

\newpage

{\large \bf Figure Caption}

\vspace{2 cm}

{\bf Fig.1}~~The general dependence of the unification scale $M_G$
(dashed line) and HSF scale $M_{HSF}$ (solid line) from the SUSY
scale $M_{SUSY}$ for the central value $\alpha _s(M_Z)=0.117$ (in
$\eta=1$ case, see Eq.(21))
\end{document}